\documentclass[12pt]{article}
\usepackage{epsfig}
\usepackage{amssymb}
\usepackage{amsmath}
\usepackage[utf8]{inputenc}
\usepackage[T1]{fontenc}
\usepackage{caption}
\usepackage{subcaption}
\usepackage{setspace}

\begin{document}

\title{A Study on the Effects of Diffusion of Information on Epidemic Spread }
\author{Semra G\"und\"u\c{c}\\
        Department of Computer Engineering \\
        Faculty of Engineering, Ankara University\\
        06345 G\"olbas{\i} Ankara, Turkey}
\maketitle

\setcounter{page}{0}

\begin{abstract}

In this work, the spread of a contagious disease on a society where the individuals may take precautions is modeled. The primary assumption is that the infected individuals transmit the infection to the susceptible members of the community through direct contact interactions. In the meantime, the susceptibles gather information from the adjacent sites which may lead to taking precautions. The SIR model is used for the diffusion of infection while the Bass equation models the information diffusion. The sociological classification of the individuals indicates that a small percentage of the population take action immediately after being informed, while the majority expect to see some real advantage of taking action. The individuals are assumed to take two different precautions. The precursory measures are getting vaccinated or trying to avoid direct contact with the neighbors. A weighted average of states of the neighbors leads to the choice of action. 

The fully connected and Scale-free Networks are employed as the underlying network of interactions. The comparison between the simple contagion diffusion and the diffusion of infection in a responsive society showed that a very limited precaution makes a considerable difference in the speed and the size of the spread of illness. Particularly highly connected hubs nodes play an essential role in the reduction of the spread of disease. 

  \vskip .2in

\noindent \rule{3in}{.01in}

\end{abstract}

\thispagestyle{empty}
\tableofcontents
\clearpage

\setcounter{section}{0}
\setcounter{page}{1}

\section{Introduction}


One of the most critical challenges of public health services is to prevent or at least control epidemic outbreaks. The process requires the optimization of the limited resources of time and supply. The resources optimization reaches its peak when society consists of well-informed individuals. There are various methods of informing the members of the society on the severity of the epidemic and the methods of prevention. Planing the methodology of the spread and increase of awareness together with an efficient vaccination program requires preliminary modeling studies.  Here in this work, a model of epidemic spread is introduced to study the effect of the awareness on the prevention of epidemic spread. The main contribution of the present work is the introduction of a mathematical relation to describe the social behavior of the individual during an epidemic spread.
The social behavior studies of Roger~\cite{Roger:2003} and the mathematical model introduced by Bass~\cite{Bass:1969} for the diffusion of innovation are taken as the base of the adoption of information and awareness spread.

$\;$

The present model aims to integrate the adoption of information with the spread of contagious diseases using aggregate and agent-based models. The spreading of infection is modeled by the well-known the susceptible-infected-removed (recovered) (SIR)  model~\cite{Kermack:1927} while the Bass model is employed for the adoption of the information.  The Bass model considers two categorically different class of individuals for the simplicity of the parametrization Bass~\cite{Bass:1969}. The first group of individuals is named as innovators. The social behavior studies of Roger~\cite{Roger:2003} indicate that the innovators respond as soon as they are informed.  In the epidemic spreading case the innovators are the individuals who adopt the idea of taking precaution as soon as they are informed. The second group is the individuals respond after observing the benefits of the new situation. The members of this group are called imitators. The imitators wait to see the decisions of the neighbors to get vaccinated of taking precaution. The majority of the individuals in a society are imitators. The information on the spread of contagious disease is formulated by using the Bass model dynamics while dynamics of the spread of the disease is governed by the well known  SIR model~\cite{Kermack:1927}. The SIR model in its original form is an aggregate model. The members of the society are classified into three groups: Susceptibles ($S$), infected ($I$) and recovered ($R$). Three coupled first order differential equations describe the dynamics of the time variation of the number of $S,\;I$ and $R$ type individuals~\cite{Kermack:1927}.  SIR model cannot accommodate intellectual individuals who can respond to incoming flux of information. For this purpose, the Bass equation added as the fourth differential equation; hence four coupled differential equations are used for the realization of concurrent diffusion of knowledge and disease.

$\;$

The agent-based model of the effects of awareness on the spread of contagious disease is introduced.
In the agent-based approach, the members of the society (agents) live on the nodes of a network. The simples network is the fully connected network in which every node is connected with every other node - everyone knows everyone. Assuming all nodes are equal (uniformity assumption), the fully connected agent-based models correspond to aggregate models. In large societies, every individual can only interact with a limited number of acquaintances. This social fact has opened possibilities of different network topologies~\cite{Castellano:2009}. One of the most commonly observed real-world network topology is the scale-free networks. In the scale-free networks, every node has a finite but varying number of connections. A small number of nodes are connected with a very high number of other nodes while the majority of the nodes have only a limited number of connections. These highly connected nodes play the hub role, connecting large parts of the network.    Information diffusion and spreading of the infection modeled on both fully connected and scale-free networks. The fully connected systems give a similar result to the aggregate model. The existence of hub nodes, in the scale-free network case considerably change the situation. It is shown that the awareness level of the individuals with a high number of neighbors is capable of stopping the disease spread. Putting the Individuals in the center of decision-making mechanism, enable the innovator group to defend themselves immediately but at the same time, the imitators make their decision according to the states of their neighbors. In both cases, the highly connected nodes naturally contribute to the prevention of the infection spreading. Comparison of the same society with different connection topologies the effectiveness of the natural prevention strategies about the high is studied.

$\;$

Although the vaccination is considered as the crucial protection strategy, other prevention methods are usually preferred by the individuals for uncomplicated virus infections. The individuals start to protect themselves almost synchronously with the information spread about the disease. Hence the information and disease spread are tightly related to the prevention of epidemics, which is the primary motivation for the present work.  The proposed model is based on two inter-woven spreading mechanisms. First one is the spread of disease and the second one is the spread of the information among the members of the society. Individual awareness and reaction play an essential role in combining these two effects. Although a wide range of studies exists (Section \ref{relatedWork}), none of them has connected sociological behavior classifications of individuals~\cite{Roger:2003} with the prevention of disease spreading.  The protection effort is considered as the natural reaction of the members of the society. This mechanism distinguishes the introduced model from the models where the source of the prevention mechanism is independent of the interactions among the members of the community.

$\;$

The work is organized as follows: The following section is devoted to outlining the related publications, the aggregate model of diffusion of contagious disease is discussed in the third section. This section mainly essential to set the rules of an agent-based model of the epidemic on networks which is presented in the fourth section. Simulation results of the agent-based model are given in the fifth section. The final section is reserved for the discussions of the results and conclusions.

\section{Related Work\label{relatedWork}}

Mathematical modeling of contagious disease spreading has many main lines of research among which, i) Estimating the model parameters from the existing data~\cite{Venkatramanan:2018,Pell:2018}, ii) resources opptimization~\cite{Zhou:2012,Yu:2018,Abouelkheir:2018}, and iii) building models of artificial societies through which the contagious diseases spread~\cite{Wang:2015} attracts considerable attention. The present work aims to address the question of the relationship between social behavior and the spreading the infection among the members of society.

$\;$

The classical epidemic models, such as the SIR model assumes a homogeneous society. The society consists of identical individuals who interact with each other with the same probability.  Societies are not uniform and homogeneous. Every member of society has a characteristic social behavior. Moreover, the interaction patterns among individuals are not uniform. Recently patterns of non-uniformity of the social connections attracted attention. A better understanding of the connectivity patterns of social networks leads to better models of social phenomena. In particular, the epidemic models became more accurate and realistic with a better understanding of the social networks~\cite{Castellano:2009} in terms of complex network theory.
Moreover, the complex-network analysis methods enabled analyzing the dynamics of the transmission of the contiguous disease more precisely~\cite{Pastor-Satorras:2001, Pastor-Satorras:2001b, Newman:2002, Wang:2016, Chowell:2016}.  The scale-free Networks which cover the majority of social networks contain some nodes with very high connectivity. The nodes with a large number of neighbors play the crucial role~\cite{Barabasi:2002} in spreading the infections.  The infection paths are cut by immunizing the highly connected nodes, which stops the spreading of the disease. The scale-free networks are robust against random attacks~\cite{Albert:2000}. Hence, random vaccination cannot stop the spread of contagious disease~\cite{Pastor:2002}. Various groups pursue the idea of immunizing individuals with a high number of neighbors.  Pastor and Vespignani proposed the Targeted immunization method which aims to protect the society from epidemic behavior by vaccination of highly connected nodes~\cite{Pastor:2002}. The high-risk immunization~\cite{Nian:2010}, based on the protection of high-risk individuals to optimize the number of nodes to be feasibly immunized. Also, Cohen et al. proposed another type of immunization strategy called acquaintance immunization: A randomly chosen neighbor of a random node is vaccinated~ \cite{Cohen:2003}.

All of the methods mentioned above require prior knowledge of the social network topology. Moreover, outside intervention is necessary to organize the vaccination of the high degree nodes. Despite all these developments, human behavior element is only recently has attracted attention~\cite{Ferguson:2007, Funk:2010}.  Modern societies are self-organizing systems. The actions of the individuals evolve to collective phenomena. Unless it is a state of emergency, very little outside control is necessary. The members of society take their actions and react accordingly.  Susceptibles may avoid contact with the infected, as a precaution, by changing their connections~\cite{Gross:2006}. The relation of the judgment of an epidemic and the behavioral changes are studied in the context of H1N1 pandemic influenza appeared in Italy in 2009~\cite{Poletti:2011}. It is shown that the individual consciousness reduces the possibility of an uncontrolled spread of a contegion~\cite{Wu:2012, Baba:2018}. Ruan et all ~\cite{Ruan:2012} proposed prevention strategies which aim to increase the awareness of the individuals.  Different behavioral groups exhibit different reactions concerning the judgment of an epidemic~\cite{Perra:2011, Xia:2014}. In opinion formation or consensus studies, the role of zealots is shown to be
very important. Similarly, the decision on taking necessary procedures against a contagion vaccination is also an opinion formation process.  The role and the effect of the committed individuals on immunization has also been discussed~\cite{Liu:2012}.

$\;$

Next section is devoted to the introduction of the proposed model of SIR model with the awareness spread. 


\section{SIR Model with Element of Information Diffusion- An Agragate approach}

The SIR model is based on an artificial society of $N$ individuals. The individuals are in three categories: susceptible ($S$), infective ($I$) and removed ($R$) where, $N = S + I + R$. The SIR model assumes a uniform society in which all individuals are in interaction with all others. Moreover, the uniformity assumption extends also to the probabilities of the transmission and recovery rates. Hence, the transmission and recovery rates of the disease are the same for all members of the society.   The SIR model has two free parameters apart from the size of the society. The first parameter, $\beta$, represents an average rate of encounters between the infected and susceptible individuals. The second parameter, $\gamma$, represents the percentage of recovery per unit time. Infected individuals recover from the illness and gain immunity. The fractions of the susceptible, infected and removed individuals can be written as, $s = S/N$, $i = I/N$, and $ r = R/N$ respectively,

\begin{eqnarray}
\label{SIRModel}
\frac{ds}{dt} &=& - \beta \, i\, s \nonumber \\
\frac{di}{dt} &=&  \beta \, i\, s - \gamma\, i  \\
\frac{dr}{dt} &=& \gamma \, i\nonumber \\
\end{eqnarray}

where $s(t)$, $i(t)$, and $r(t)$ are the fractions of susceptible, infected and removed individuals at a given time $t$.

$\;$

Although the uniformity assumption is an oversimplification of the situation, the model still successfully explains the spreading process of contagious diseases.  The first and most important prediction of the model is the existence of a threshold value which is a function of transmission and recovery rates. The model can also accommodate various assumptions for immunization and vaccination.

$\;$

In real societies, as soon as an infection starts to spread, information on the spread rate and level of Infectiousness becomes available to the public through two channels: Public broadcasting (official data) and word-of-mouth (in technology-oriented societies, social media).  One way of integration of information diffusion and the SIR model requires reinterpretation of the Bass model. The number of individuals who adopt information, at time $t$ is $V(t)$; which leads to the percentage of vaccinated individuals, $v(t)=V(t)/N$.  The modified Bass equation becomes,

\begin{equation}
  \label{BassModel}
\frac{dv(t)}{dt} = (p + q (v(t) + i(t)) + r(t) ) (1 -  (v(t)+i(t)+r(t))
\end{equation}

where $v(t),i(t)$ and $r(t)$ are the percentages of vaccinated, infected and recovered individuals.  $s(t) = (1 - (v(t)+i(t)+r(t))$ gives the percentage of susceptibles who are open to infection.  The parameters $p$ and $q$ are known to be innovation and imitation parameters of the Bass model. Here, the innovation parameter best understood as the probability of adoption the new situation (get vaccinated) immediately after being informed. The imitations parameter is the probability of adoption after having some experience through the neighbors (Word-of-mouth). Imitators try to protect themselves after seeing the benefit of vaccination through vaccinated, recovered or infected neighbors.

$\;$

Coupling SIR model equations with the Bass equation leads to modeling
a society with individuals who may take precursory measures under the
spread of contagious disease~\ref{ModifiedSIR}.

\begin{eqnarray}
\label{ModifiedSIR}
\frac{ds}{dt} &=& - \beta \, i\, s - (p + q ( v +  i) ) s\nonumber \\
\frac{di}{dt} &=&  \beta \, i\, s - \gamma\, i  \\
\frac{dr}{dt} &=& \gamma \, i\nonumber \\
\frac{dv}{dt} &=& (p + q ( v +  i) ) s.\nonumber
\end{eqnarray}

In these four coupled first order differential equations, susceptible individuals become infected with the rate $\beta$, infected individuals recover with the rate of $\gamma$. At the same time some susceptible $s = 1 - (v + i + r) $, take precursory measures and become immune. The parameter $p$ and $q$, determine the vaccination rates of the susceptibles who adopt the idea of vaccination immediately after being informed and after observing the benefit respectively. In this work, the probability of transmission $\beta=1.0$ and the probability of recovery $\gamma=0.2$ are kept constant for both aggregate and the agent-based models.

$\;$

\begin{figure}[h]
    \centering
     \begin{subfigure}[b]{150pt}
        \includegraphics[width=\textwidth]{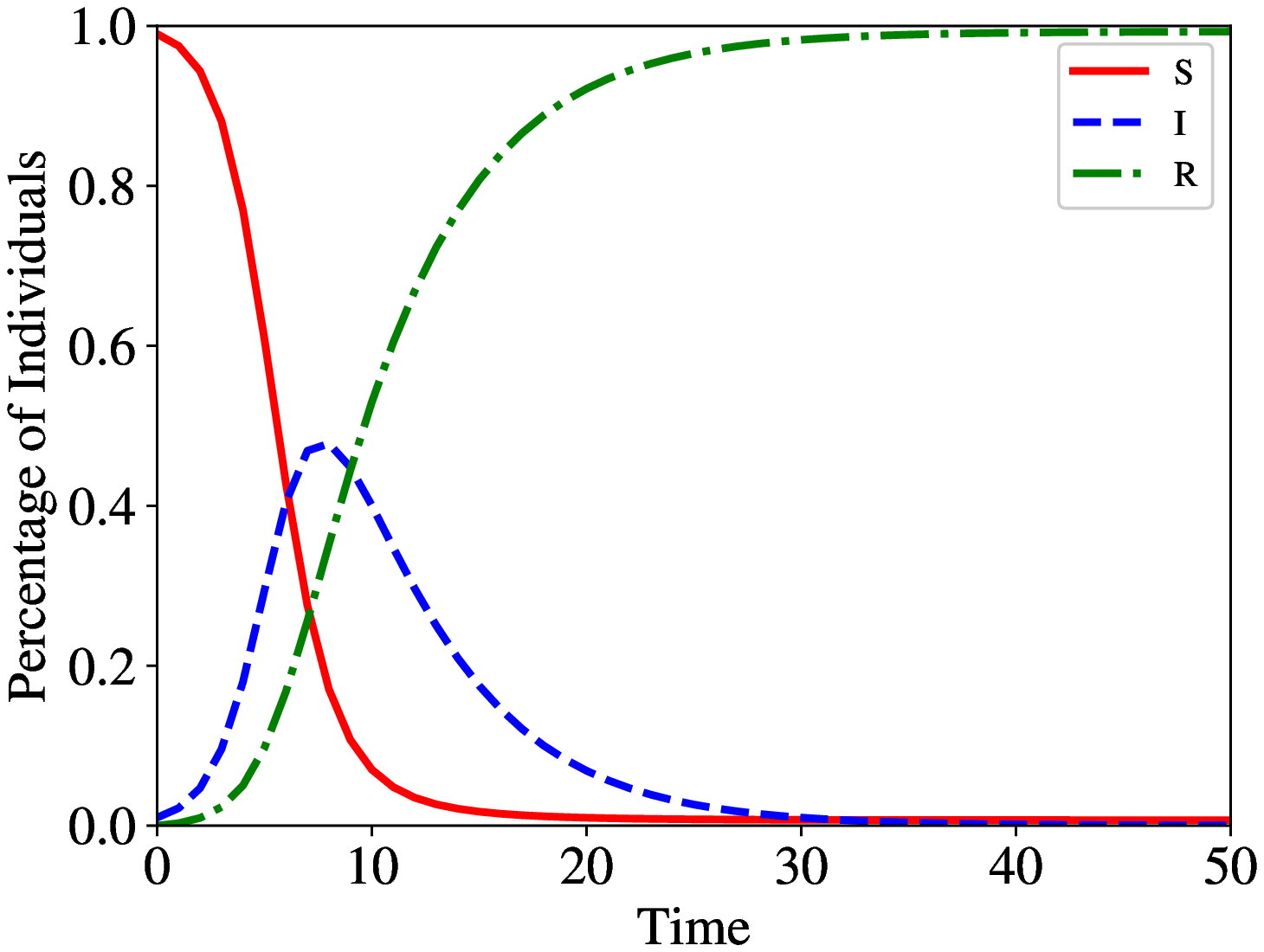}
        \caption{SIR Model}
        \label{fig:SIR}
    \end{subfigure}
    \begin{subfigure}[b]{150pt}
        \includegraphics[width=\textwidth]{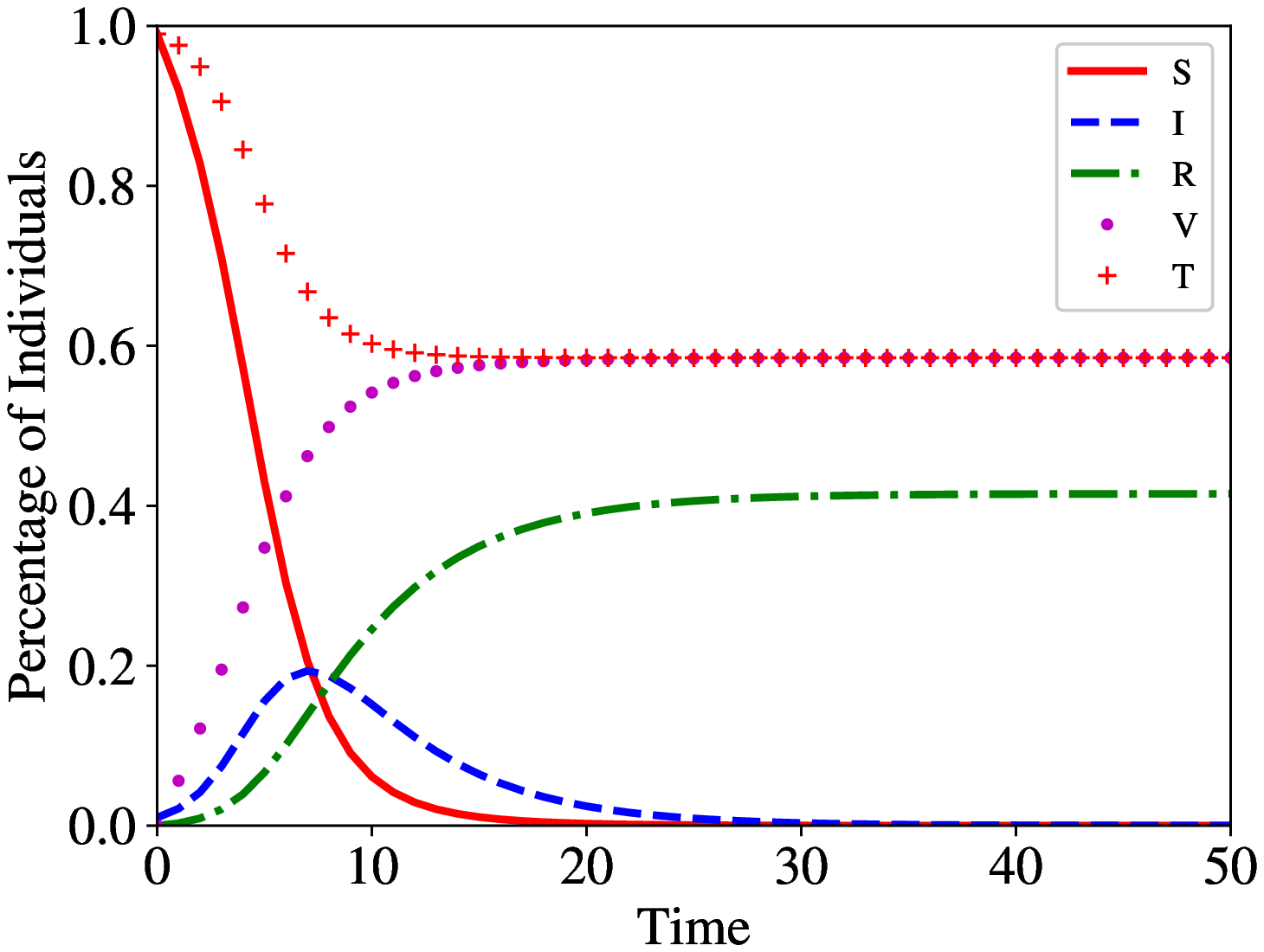}
        \caption{SIR model with vaxination}
        \label{fig:ExtendedSIR}
    \end{subfigure}  
    \caption{Spread of epidemics in a uniform society.  }\label{fig:SIR&ExtendedSIR}
\end{figure}

Figures \ref{fig:SIR} and \ref{fig:ExtendedSIR} show the effect of opinion diffusion among the members of society.  The difference between the figures \ref{fig:SIR} and \ref{fig:ExtendedSIR} is the number of individuals who can avoid contamination.

For the information spread, only virtual interactions are sufficient. The spread of contamination requires direct contact interaction. In this sense even if the social network is the same since information and contamination spread are different types of interactions with different rules, their spread dynamics are different. The information of the spread of the infection reaches the susceptible through the infected, recovered and vaccinated neighbors.  Upon this information the susceptible makes a decision: Take a precursory measure or not.  In the decision-making process,  both infected and vaccinated individuals contribute equally (Equation \ref{ModifiedSIR}).  Figure~\ref{fig:SIR} show SIR equation (Eqn. \ref{SIRModel}) solution with the above-given parameters. The figure shows the expected susceptible-infective-recovered relation. The chosen parameters are higher than the threshold value of the epidemic; hence all of the members of the population becomes infected, and in time they recover. In $50$ discrete time steps, the population consists only of the recovered individuals.  Figure \ref{fig:ExtendedSIR} shows the results of the extended SIR model (Eq. \ref{ModifiedSIR}).  The extension, Bass equation as the fourth differential equation requires the innovation and imitation parameters.  From the marketing studies, it is shown that the innovation parameter is much smaller than the imitation parameter.  This situation is intuitively understandable since the number of individuals who take risks is less than the imitators, who follow the trend or observe the benefits of the new
situation, the innovation parameter is less than the imitation parameter.  The innovation and imitation parameter values, $p = 0.05$ and $q = 0.2 $ are also used for both aggregate and agent-based models. In the SIR model case, the total number of infected individuals summed up to $100 \%$ of the population. The voluntary vaccination/protection program with the above-given parameters changes the situation. The innovation and imitation parameter values,$p = 0.05$ and $q=0.2 $, reduce the peak of the infected individuals down to $20\%$ level. With such low infected ratio, the recovered individuals remain around $40\%$, while $60\%$ of the population remains unaffected by the infection.


Such a totalistic approach cannot produce the expected result of the intelligent behavior. The characteristics of the real social networks implemented in the above-described model. The following section has devoted the application of the aggregate model on the different network topologies to discuss the effects of the network topologies on the spread of contagious disease under the intelligent selective
vaccinations.

\section{Rules of Agent Based Model on Networks}


For agent-based simulations, the model society a collection of interacting units (agents). The interactions rules and the network of connections of the agents determine the dynamics of the evolving social phenomena. The network contains $N$ nodes.  Each node, $N_i$, carries a state variable, $S,\; I,\; R$ or $V$ for susceptible, infected, recovered and vaccinated states respectively. At the initial stage all nodes have common infection, $\beta_i=\beta$ and recovery $\gamma_i=\gamma$ probabilities. After the beginning of the spread of infection, the collected information leads some nodes to take precursory measures.  The uniformity of the fully connected networks did not allow individuals to make a choice. On the contrary with the fully connected network case, here, agents can choose to reduce the risk of being contaminated or choose to be vaccinated.

The states of the agents changed through mutual interactions. The number of reciprocal interactions is the measure of the discrete time steps. The spread of infection is followed in time steps until the stationary state is reached.

$\;$

At one time step, the states of $N$ randomly selected nodes are sequentially updated. Update of a node is the dynamical change through interaction with a randomly chosen neighbor. Interaction of two connected sites sets their state for the next time step. The rules of transition are as follows, a)If a susceptible node interacts with an infected neighbor, they mutually contaminate each other. b) If the chosen node is infected, may recover with probability $\gamma$. c) If the node is susceptible and did not meet an infected neighbor until the time of update, get vaccinated immediately with the transition rate $p$ or gather information on the number of infected, recovered and vaccinated neighbors. If the collected data triggers an action, gets immunized with the probability $q_1$.  The node changes its state from $S$ to $V$, and this node gains immunity. Another precursor measure is to reduce the probability of neighbor induced contamination. The node who take precursory action reduce its transition probability with probability $q_2$, avoids infection, remain susceptible.

$\;$

Two separate sets of rules are necessary for the diffusion of contamination and information gathering.  The first set of rules are related to the contact interaction of the neighboring agents. The proposed rules for transmission of the contiguous disease on networks have the following steps:

\begin{enumerate}
\item A randomly selected individual ($N_i$) interacts with a randomly selected neighbor ($N_j$).
\item If either one of them is infective, at time slice $t$  while its neighbor is susceptible, the susceptible the partner becomes infected with a probability $\beta$ at the time slice $t+1$
\item If the selected individual ($N_i$) is already infected, recovers with probability $\gamma$ at time slice $t+1$.
\end{enumerate}

At each time step, all nodes mutually exchange information.  The information sent by every node contains the state of the node. Each node evaluates the collected data.  If the individual $N_i$ is susceptible at time step $t$ decides whether to take precursor or not. The rules of information gathering and decision-making process is as follows:

If an individual living at vertex $i$ is susceptible, at time $t$,

\begin{enumerate}
\item A node $N_i$  (innovator) gets vaccinated with probability $p$, regardless of the state of the neighbors. 
  
\item If the individual is an imitator, take precursory measures with the probability proportional with the states of the neighbours,
\[ P_{Neighbors} =  \;\frac{\sum_{\alpha \in N\\N} k_{\alpha} \left( \delta_{N_{\alpha},I}+\delta_{N_{\alpha},V}+\delta_{N_{\alpha},R}\right) }{\sum_{\alpha \in NN} k_{\alpha}} \].
If $P_{Neighbors} r$ the individual living at site $i$ becomes aware of the danger and take action.
Here, $r$ is  a  uniform random number, $0\le r < 1$.  For an imitator there are two possible choices:
\begin{enumerate}
\item  With probability $q_1$ get vaccinated and gain immunity, 
\item with probability  $q_2$ take measures and reduce the probability of being infected.
\end{enumerate}
 Any action other than vaccination does not give full immunity but reduce the probability of infection  $\beta_i \rightarrow \beta^{\prime}_i$.  In this study, $\beta^{\prime}_i$ is taken as $\beta_i/4$.

\[
\begin{array}{lll}
{\rm if}&\;\; q_1 > r_1 &\;\;\; N_{i} = V\\
{\rm else\; if}&\;\; q_2 > r_2 &\;\;\; \beta_{i} = \beta_i/4\\
{\rm else}     &               &{\rm does \; noting.} 
\end{array}
\]

Here $r_1$ and $r_2$ are random numbers with uniform distribution, $q_1$ and $q_2$ are imitation parameters for vaccination and other protective measures respectively. Initially, all vertices are assigned the same value of infection probability value. The infection probability is reduced to one-fourth for the individuals who choose the protection as a measure of protection.
\end{enumerate}

In the following subsection, the model will be tested using scale-free networks will be tested to observe the effects of topology differences of the social networks.

\section{Agent Based Model on Scale Free Networks}


One of the most prominent examples of real-life networks is the scale-free networks. The Barab\'asi-Albert~\cite{Barabasi:1999} preferential-attachment algorithm is one of the commonly used algorithms to create scale-free networks. In this work, the  Barabasi-Albert algorithm is employed to create scale-free networks.
The network starts with $m$ seed nodes which are mutually connected. Each new node makes $m$ links to the existing nodes with probability, \[P = \frac{k_i}{\sum_j k_j}\] where $k_i$ is the degree of the $i^{\rm th}$ node. The sum of the degrees of all existing nodes is the normalization term. The number of the initial nodes plays an important role in the topology of the network. If the network starts with only very few numbers of nodes, some of the nodes will be highly connected.  The highly connected nodes are called hub nodes since such nodes transmit the effect to a very high number of nodes. This situation is typical of the scale-free networks. As the number of initial nodes increases, the number of hub nodes increase but their connectivity decrease. If the network starts with a high number of nodes, none of the nodes will have a very high degree, and the network resembles random networks.

$\;$

\begin{figure}[ht]
    \centering
     \begin{subfigure}[b]{0.45\textwidth}
        \includegraphics[width=\textwidth]{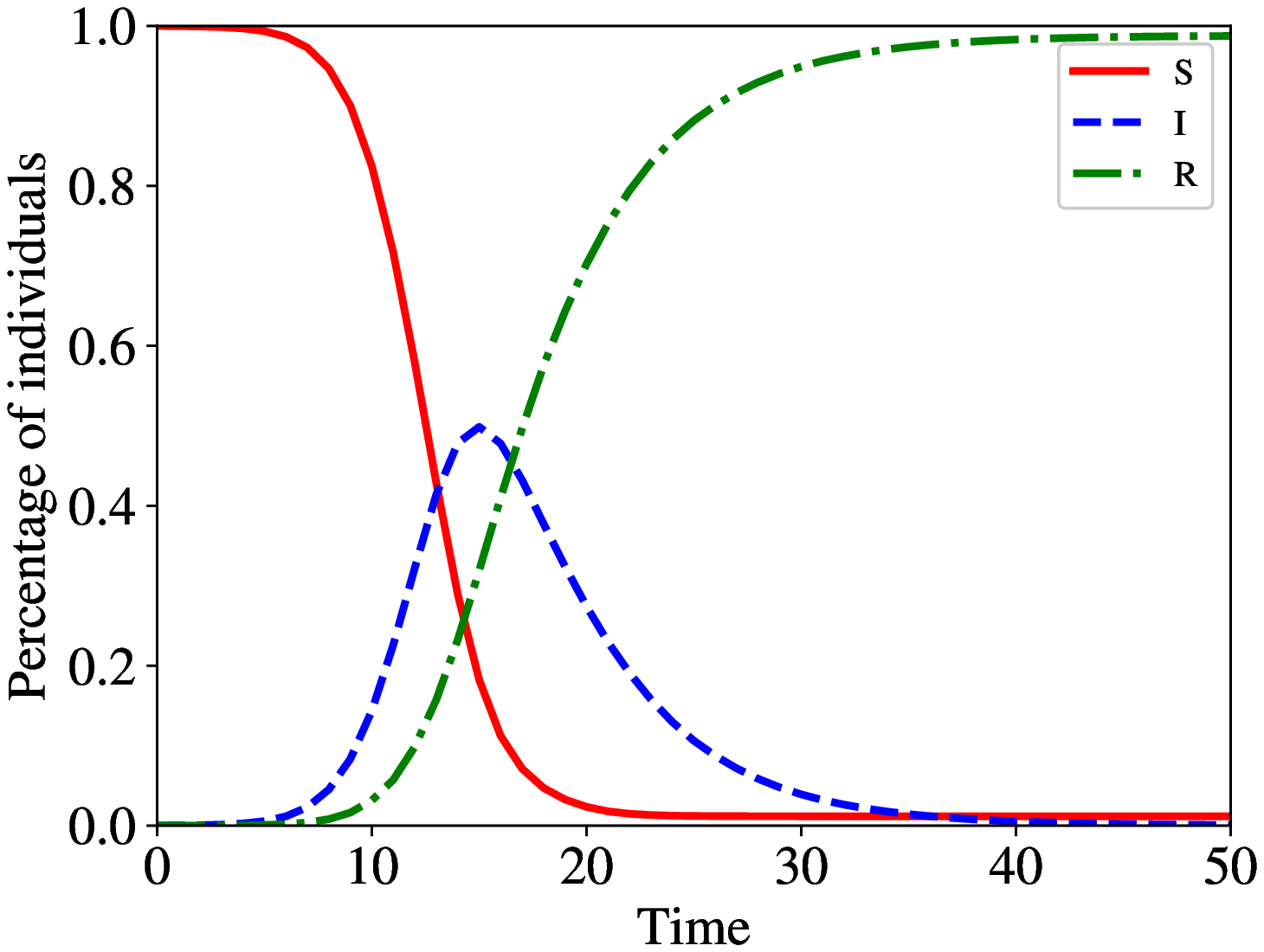}
        \caption{SIR model on scale-free network ( $m=2$)}
        \label{fig:SFM02SIR}
    \end{subfigure}
    \begin{subfigure}[b]{0.45\textwidth}
        \includegraphics[width=\textwidth]{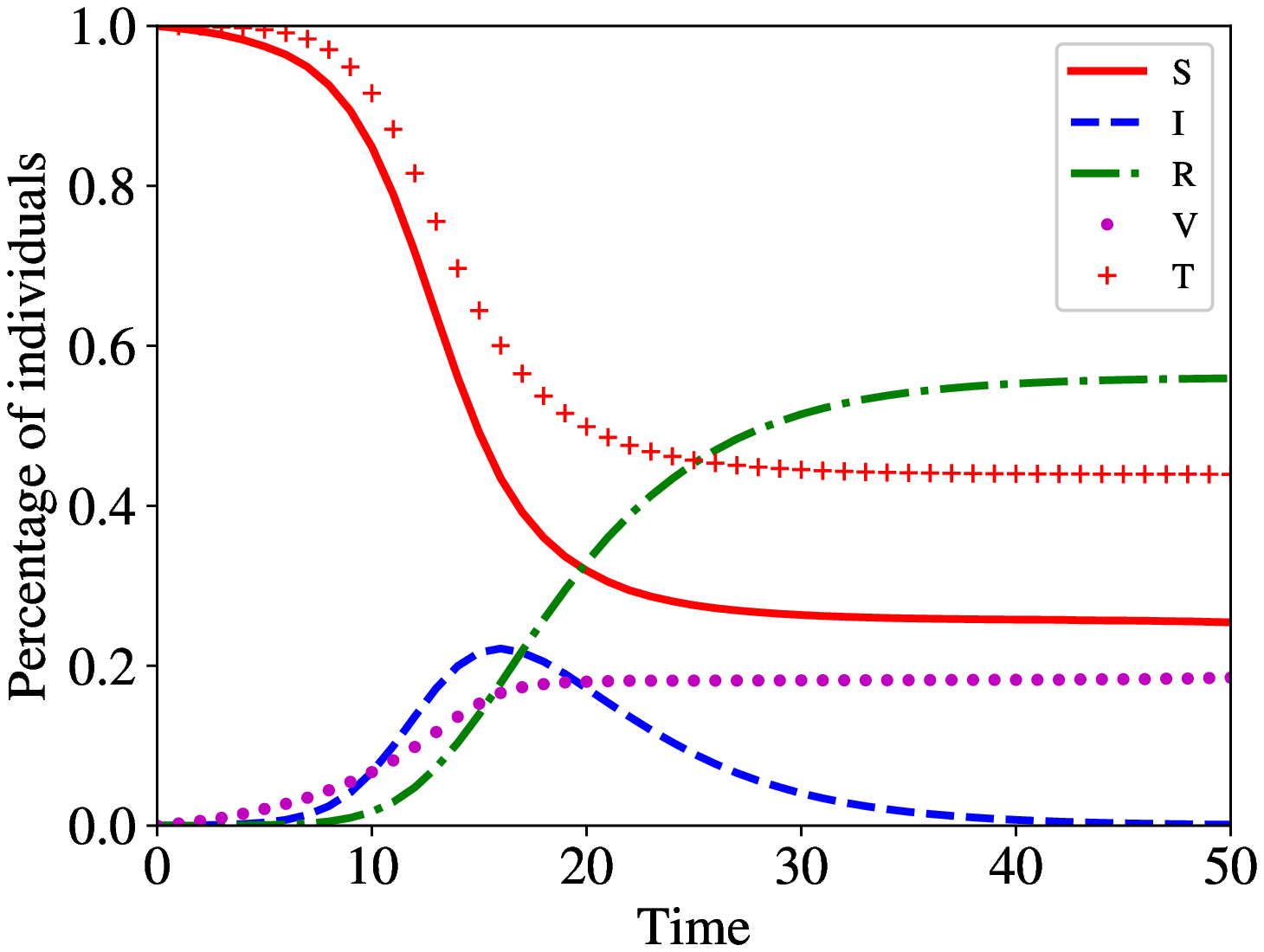}
        \caption{SIR with vaxination on scale-free network ($m=2$)}
        \label{fig:SFM02Vaxination}
    \end{subfigure}  \\
     \begin{subfigure}[b]{0.45\textwidth}
        \includegraphics[width=\textwidth]{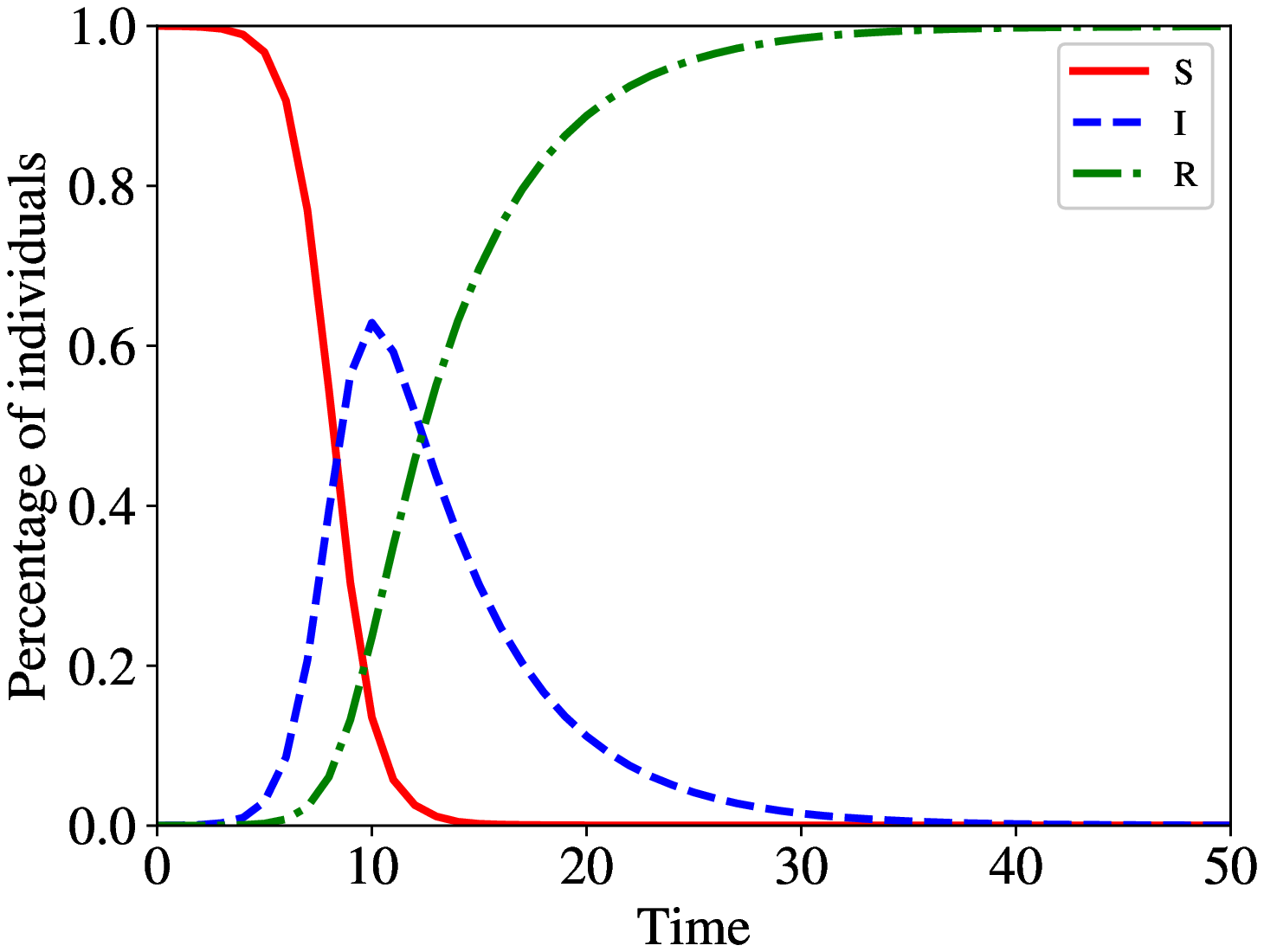}
        \caption{SIR on scale-free network ($m=20$)}
        \label{fig:SFM020SIR}
    \end{subfigure}
    \begin{subfigure}[b]{0.45\textwidth}
        \includegraphics[width=\textwidth]{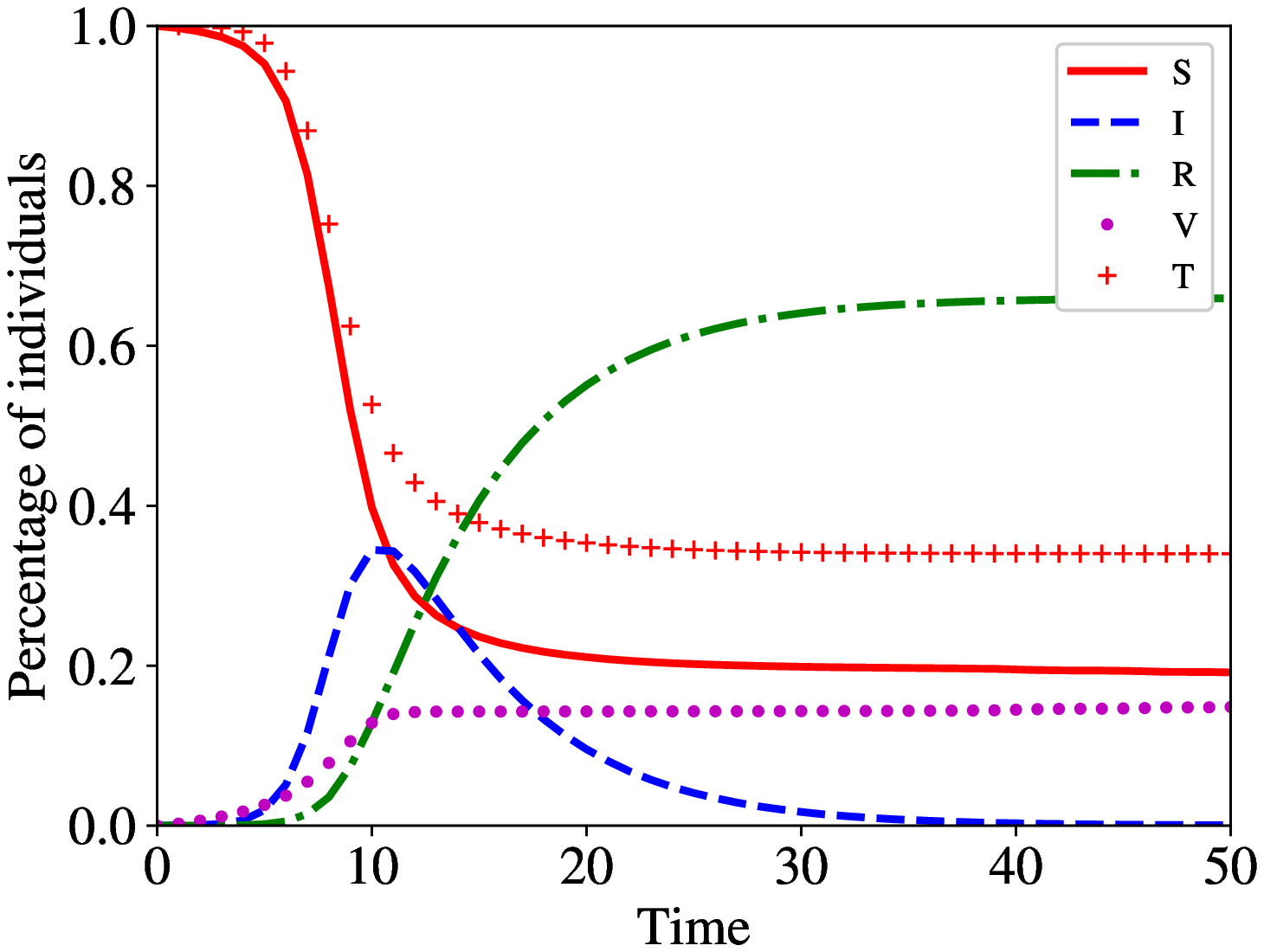}
        \caption{SIR with vaxination on scale-free network ($m=20$)}
        \label{fig:SFM020Vaxination}
    \end{subfigure}  
    \caption{Spread of contegious disiese in a society with scale-free network topology.}\label{fig:ScaleFree}
\end{figure}

The spread of contagious disease on scale-free networks are summarised in Figure \ref{fig:ScaleFree}. The model society consists of $10000$ individuals. A scale-free network provides the connections among the members of the community.  Networks with the number of seed nodes, $m=2,$ and $m=20$ are used to study the role of highly connected nodes. Highly connected hub nodes are the characteristics of the scale-free network with a small number of initial states. The increasing number of seed nodes change the structure of the network. As the number of initial nodes increases, the hub nodes lose their importance.  With an increasing number of seed nodes the scale-free networks gradually divert to random networks.

$\;$ Initially, all nodes are in the susceptible state, $S$. One of the randomly selected nodes is assigned as infected $I$. Measurements are done over $100$ statistically independent initial configurations and new sets of connections. Each configuration contained only one infected individual.  The time evolution of each new system is followed $50$ time steps.

$\;$
Each system contains two elements: The network and the distribution of the states of the nodes. Each node is assigned transmission, and recovery rates as well as the innovation, imitation parameters. Initially, the parameter values are uniformly distributed to all nodes. During the simulation, only the infection transmission rate $\beta$ is allowed to change through interactions. The same network is employed for both spreads of contamination and information. Figures \ref{fig:SFM02SIR} and \ref{fig:SFM020SIR} show the spread of contamination in two, same size, networks. The difference between these networks is in their connectivity structure.  The first network centered around $2$ seed nodes while the second network started with  $20$ seed nodes. Hence, the degree distributions of highly connected nodes are different.  Just a few highly connected nodes exist in the $m=2$ case, while $m=20$ is more like a random network.  In both cases, the parameters related with the transmission of illness ($\beta$ and $\gamma$), and parameters related with the adoption of action ($p$,$q_1$ and $q_2$) parameters are the same for both networks. Moreover, the fixed values of the probabilities, $\beta = 1$ and $\gamma=0.2$ are the same as in the aggregate model case. The difference is due to the change of the network topology.  The number of infected individuals is higher in the network with a large number of initial sites.  Moreover, the peak value of the number of infected is reached before the peak of the network originated from only a small number of initial nodes. This is understandable since the connectivity of the random networks spread of the infection equally well,  regardless of the choice of the position of the initial infected individual. In the stong hub case, a limited number of initial nodes have very high connectivity, but rest of the nodes are connected with only a small number of neighbors. Figures \ref{fig:SFM02Vaxination} and \ref{fig:SFM020Vaxination} show the effects of the individual prevention efforts for the networks initiated with $m=2$ and $m=20$ nodes .

$\;$

The case of the societies with intelligent individuals who take precaution when the disease information spread is presented by the figures \ref{fig:SFM02Vaxination} and \ref{fig:SFM020Vaxination}.  The figures include the changes in the number of susceptibles $S$, infected, $I$ recovered $R$, vaccinated $V$ and the total number of non-contaminated individuals, $T$. Vaccinated or precautioned hub nodes reduce the probability of being contaminated blocks the paths of contamination. In this sense, the hub nodes play an important role in reducing the speed and the size of the spread which also enables the other to take precaution. This effect is presented by the figures, \ref{fig:SFM02SIR} and \ref{fig:SFM02Vaxination}. Figure \ref{fig:SFM02SIR} show that at the peak, the infected ratio reaches about $60 \%$ of the population. The contamination spreads to all members of the society at the end of $50$-time slices.  Figure \ref{fig:SFM02SIR} show that, individual prevention efforts are very efficient in the case of $m=2$. The peak value of the infected individuals hardly reaches one-fifth of the population. After $50$ time steps, slightly less than half of the population, $T$, remain unaffected by the disease.


Using the fixed infection transition and recovery rates, the spread of the infection is faster on networks with a higher number of initial nodes ($m=20$) than the small number of seed nodes ($m=2$).  In the case of a high number of seed nodes also the peak of the infected individuals reach the point which is higher than the half of the population (Figure~\ref{fig:SFM020SIR}).  The majority of the
population has already been infected after $10$ time steps.  Since the peak is reached at about ten iterations, individuals have less time to take precaution. The number of vaccinated remains less than the case of $m=2$. Therefore the total number of unaffected individuals are also considerably less than $m=2$ case.

\begin{figure}[ht]
    \centering
     \begin{subfigure}[b]{0.45\textwidth}
        \includegraphics[width=\textwidth]{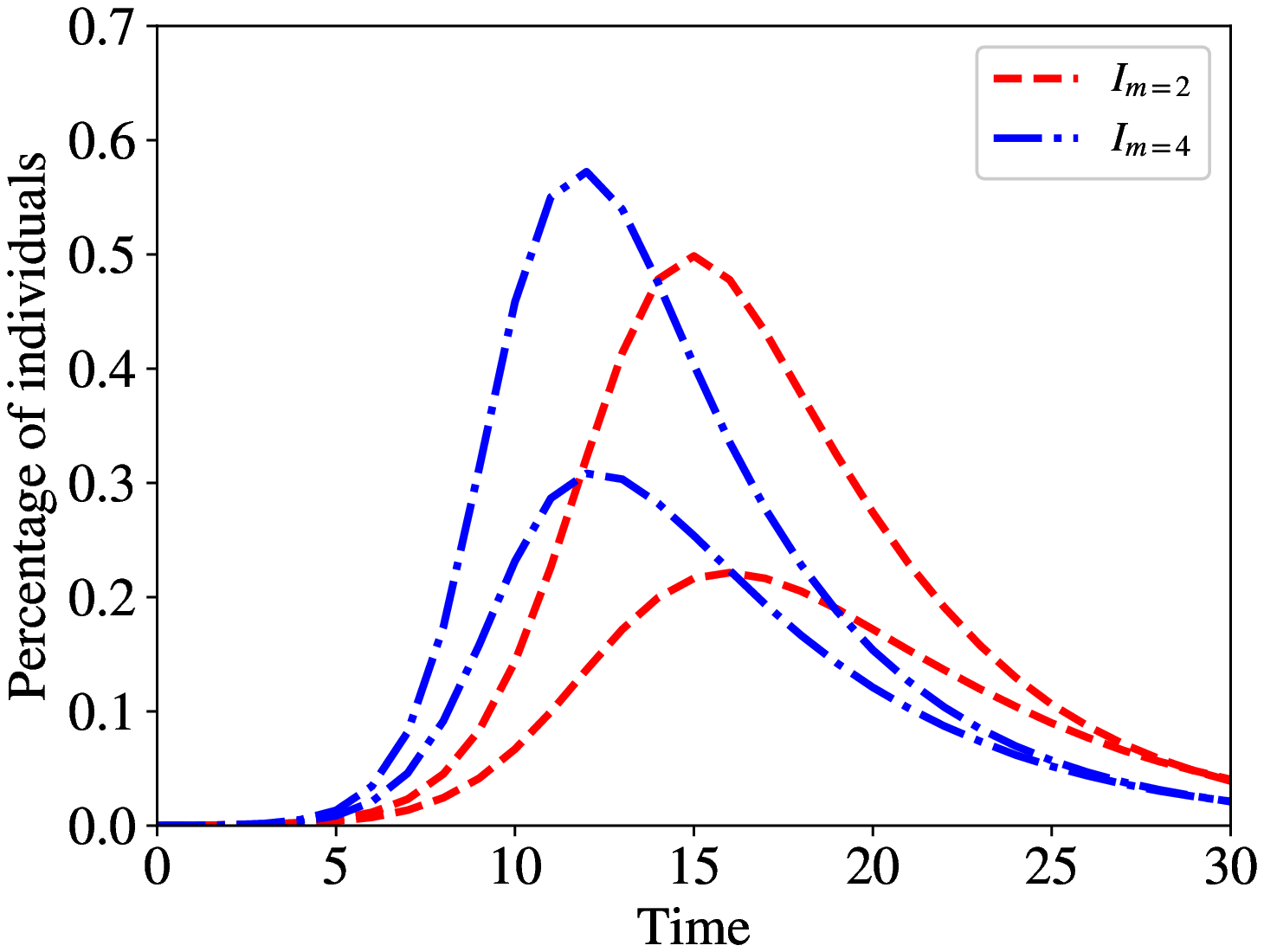}
        \caption{$m=2$ and $m=4$ }
        \label{fig:SF_Small}
    \end{subfigure} 
    \;
     \begin{subfigure}[b]{0.45\textwidth}
        \includegraphics[width=\textwidth]{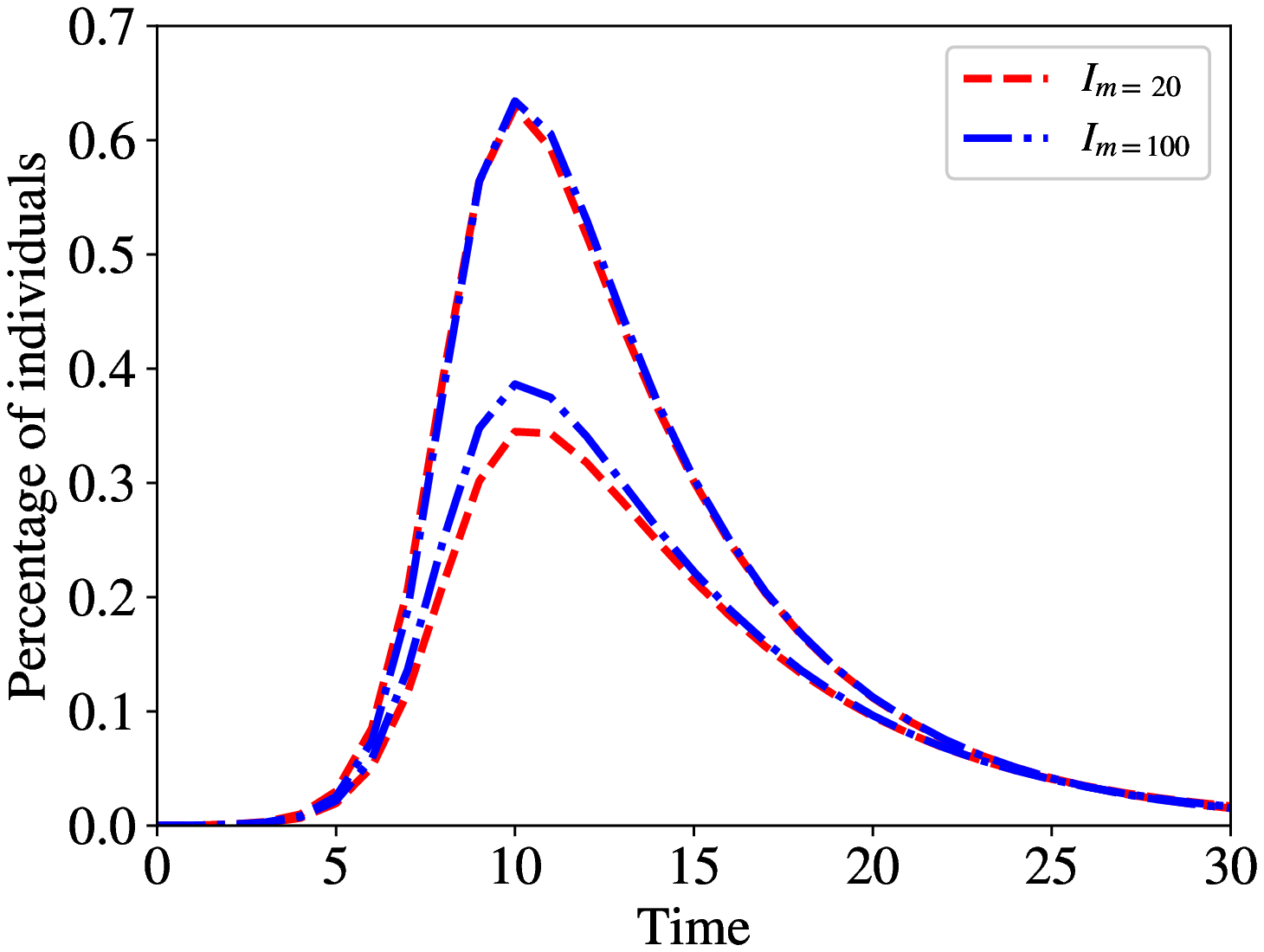}
        \caption{ $m=20$ and $m=100$ }
        \label{fig:SF_Large}
    \end{subfigure} 
    \caption{The effects of information processing on the  number of infected individuals. a) $m=2$ and $m=4$, b) $m=20$ and $m=100$ . }\label{fig:PeakComparisson}
\end{figure}


Figure \ref{fig:PeakComparisson}, discuss the role of connectivity in
the spread of infection. Figure \ref{fig:SF_Small} and
\ref{fig:SF_Large} show the spread of disease in scale-free networks
with a low and high number of hub nodes. Figure \ref{fig:SF_Small} exhibits the effect of the changes in the number of highly connected
hub nodes on the number of infected individuals. The Figure
\ref{fig:SF_Small} show the comparison between $m=2$ and $m=4$ networks.  In the case where highly connected hub nodes exist ($m=2,4$) the effects of the individual precaution attempts to reduce the
initial high peak value of the infected individuals.  As the number of
hub nodes increases, the diffusion speed increases, consequently the
prevention efforts are not sufficient. The Figure \ref{fig:SF_Large} is devoted to the comparison of a high number of hub nodes, $m=20$ and
$m=100$, where it is seen that as the number of hub nodes increase, a
limiting link distribution is reached. At this point, the diffusion
speed of the infection reaches its peak.


\section{Discussions and Conclusions}

 In this work, the aim is to model the spreading of disease on networks while susceptibles exchange information on the state of the neighbors. The accumulated knowledge is used to make decisions on the possible precautions. The infection spreads among the members of the community through direct contact interactions while the information exchange with the neighbor sites is simultaneous and does not require direct contact. The susceptible members of society may take measures to prevent infection. The proposed methods of prevention are being vaccinated or trying to avoid direct contact with the neighbors. The collected information on the states of the neighbors leads to the choice of precursory action.

$\;$

The fully connected and scale-free networks are used to test the proposed model.  On the fully connected system, the comparison between the SIR model and the proposed model with awareness shows a considerable reduction in the spread rate of infection. Even a limited number of precaution taken nodes are sufficient for such significant success in fighting epidemics.  In the scale-free network case, the same observations valid and moreover it is seen that the highly connected nodes are essential for the control of the spread of disease. A precautioned hub node can block a high number of connections, reducing the spread rate of infection.

$\;$

All types of networks are under attack of different viruses, computer viruses, and malice information.  In real-life situations, nodes defend themselves from the fatal attacks. Nodes continuously exchange information.  The shared crucial information help nodes to prepare themselves for possible malice situation. This work aimed to model existing real-life protection mechanisms appearing in networked systems. The role of connection topology is seen to play a crucial role in prevention attempts. In general, contact networks are separate than the communication networks the disease transmission is of concern.  For technological networks, both communication and interactions are part of the same system. When the direct interactions and communications are on different networks,  multilayer networks must substitute traditional network structure. The transmission of information and infectious disease on multilayer networks will be the subject of future work.


\end{document}